\documentclass[sigconf]{acmart}

\AtBeginDocument{%
  }

\usepackage{booktabs}
\usepackage{multirow}
\usepackage{enumitem}
\usepackage{xcolor}
\usepackage{xspace}
\usepackage{csquotes}
\usepackage{pgfplots}
\pgfplotsset{compat=1.18}
\usepackage{subcaption}
\graphicspath{{fig/}}

\usepackage[normalem]{ulem}

\newcommand{\method}{SetMIR}
\DeclareMathOperator*{\argmax}{arg\,max}
\DeclareMathOperator*{\argmin}{arg\,min}

\setcopyright{none}
\copyrightyear{2027}
\acmYear{2027}
\acmDOI{}
\acmConference[KDD '27, ADS Track, Under Review]{The 33rd ACM SIGKDD Conference on Knowledge Discovery and Data Mining}{August 1--5, 2027}{San Jose, CA, USA}
\acmISBN{}

\begin{document}

\title[SetMIR]{SetMIR: Multi-Interest Retrieval as Set Prediction}

\author{Xiaodong Liu}
\affiliation{%
  \institution{Snap Inc.}
  \city{Bellevue}
  \state{WA}
  \country{USA}
}
\email{xliu9@snapchat.com}

\author{Congfei Zhang}
\affiliation{%
  \institution{Snap Inc.}
  \city{Bellevue}
  \state{WA}
  \country{USA}
}
\email{czhang3@snapchat.com}

\author{Hsiang-wei Chao}
\affiliation{%
  \institution{Snap Inc.}
  \city{Seattle}
  \state{WA}
  \country{USA}
}
\email{hchao@snapchat.com}

\author{Siman Wang}
\affiliation{%
  \institution{Snap Inc.}
  \city{Bellevue}
  \state{WA}
  \country{USA}
}
\email{swang7@snapchat.com}

\author{Xiao Bai}
\affiliation{%
  \institution{Snap Inc.}
  \city{Palo Alto}
  \state{CA}
  \country{USA}
}
\email{xbai@snapchat.com}

\author{Tong Zhao}
\affiliation{%
  \institution{Snap Inc.}
  \city{Bellevue}
  \state{WA}
  \country{USA}
}
\email{tzhao@snapchat.com}

\author{Jingxiao Ma}
\affiliation{%
  \institution{Snap Inc.}
  \city{Bellevue}
  \state{WA}
  \country{USA}
}
\email{jma3@snapchat.com}

\author{Wen Zhang}
\affiliation{%
  \institution{Snap Inc.}
  \city{Bellevue}
  \state{WA}
  \country{USA}
}
\email{wzhang7@snapchat.com}

\author{Zhe Liu}
\affiliation{%
  \institution{Snap Inc.}
  \city{Palo Alto}
  \state{CA}
  \country{USA}
}
\email{zliu11@snapchat.com}

\author{Shantanu Aggarwal}
\affiliation{%
  \institution{Snap Inc.}
  \city{Palo Alto}
  \state{CA}
  \country{USA}
}
\email{saggarwal@snapchat.com}

\author{Di Huang}
\affiliation{%
  \institution{Snap Inc.}
  \city{Santa Monica}
  \state{CA}
  \country{USA}
}
\email{dhuang6@snapchat.com}

\author{William Leach}
\affiliation{%
  \institution{Snap Inc.}
  \city{Seattle}
  \state{WA}
  \country{USA}
}
\email{wleach@snapchat.com}

\author{Yunzhi Zhou}
\affiliation{%
  \institution{Snap Inc.}
  \city{Palo Alto}
  \state{CA}
  \country{USA}
}
\email{yzhou10@snapchat.com}

\author{Yajun Wang}
\affiliation{%
  \institution{Snap Inc.}
  \city{Palo Alto}
  \state{CA}
  \country{USA}
}
\email{ywang30@snapchat.com}

\author{Jinchao Li}
\affiliation{%
  \institution{Snap Inc.}
  \city{Bellevue}
  \state{WA}
  \country{USA}
}
\email{jli18@snapchat.com}

\author{Yu Zhang}
\affiliation{%
  \institution{Snap Inc.}
  \city{Palo Alto}
  \state{CA}
  \country{USA}
}
\email{yzhang3@snapchat.com}

\renewcommand{\shortauthors}{Liu et al.}

\begin{abstract}
Embedding-based retrieval is at the core of industrial recommender systems, but a single user embedding is often too limited to capture a user’s diverse interests. Multi-interest retrieval addresses this by using multiple user embeddings, yet existing methods still suffer from two issues: \emph{interest collapse}, where different embeddings learn the same interest, and \emph{static dispatch}, where serving uses a fixed retrieval budget even when some embeddings are unnecessary. We propose \textbf{SetMIR}, which treats multi-interest retrieval as a set prediction problem. SetMIR encodes a user’s behavior history with a transformer and uses $K$ learnable queries to decode a set of user interests, each producing a retrieval embedding and a presence score. During training, Hungarian matching assigns targets to queries one-to-one, so matched queries learn distinct interests and the presence head learns which queries are active. At serving time, SetMIR uses presence scores and query-level Non-Maximum Suppression (NMS) to issue only active, non-redundant ANN queries. On Snap’s Dynamic Product Ads (DPA) data, SetMIR outperforms four learned multi-interest retrievers on every metric while issuing $33\%$ fewer ANN queries per request. Deployed as a new retrieval source in the DPA production stack, SetMIR lifts overall CVR by 3.1\%, while lifting CTR by $44\%$ and CVR by $51\%$ over the item-to-item retrieval source with the same item embeddings, ANN index, and retrieval quota.
\end{abstract}



\maketitle

\section{Introduction}
\label{sec:intro}

The retrieval stage of a large-scale recommender selects a few
thousand candidate items per request from a corpus of millions to
billions; a downstream ranker then scores those candidates with a
heavier model~\cite{covington2016youtubednn,huang2020facebook}. The
dominant retrieval architecture is a two-tower neural network: a user
tower produces one embedding, an item tower indexes the catalog, and
approximate-nearest-neighbor (ANN)
search~\cite{huang2013dssm,yi2019sampling,guo2020scann,johnson2021faiss}
retrieves the top-$N$ items by inner product or cosine similarity. A
single user embedding is often too coarse for this task. A user may
interact with streetwear, baby clothes, and skincare in the same week,
yet an averaged representation favors dominant categories and misses
narrower, short-lived ones.

Multi-interest retrieval addresses this limitation by producing $K$
embeddings per user, each issuing an ANN query whose results are
merged~\cite{li2019mind,cen2020comirec,fan2025dcm}. However, current
methods usually have two drawbacks. First, \textbf{interest collapse}:
the standard training signal~\cite{li2019mind} assigns each target
to its closest interest by $\argmax_k \langle e_k, t \rangle$;
multiple targets can collide on the same query, unmatched queries
receive no gradient, and the model can converge to only a few
effective interests. Second, \textbf{static dispatch}: issuing ANN
searches from \emph{all} $K$ embeddings spreads a per-request retrieval budget
$N$ across redundant or low-quality queries, and without a learned
per-query quality signal, $K$ ANN calls may return fewer than $K$
meaningfully distinct candidate sets while still consuming the full
budget. Recent methods add auxiliary
regularizers~\cite{zhang2022re4,xie2023remi}, but they do not learn
which of the $K$ embeddings should retrieve candidates for a
particular user request.

We propose \method{}, a set-based multi-interest retrieval framework
that predicts a variable-size set from a fixed-size bank of learnable
queries: the target-set size is unknown, and the queries must
coordinate so each captures a distinct element. A Hungarian matcher~\cite{kuhn1955hungarian}
assigns engaged targets to queries one-to-one, so each matched query
receives a retrieval gradient while unmatched queries receive an
explicit absence target. A per-query \emph{presence} score is supervised by the matcher
(matched vs.\ unmatched) and gates which of the $K$ queries dispatch
ANN searches at inference time. Our
contributions are summarized as follows:
\begin{itemize}
\item We frame multi-interest retrieval as set prediction over $K$
learnable queries, which removes interest collapse at the
training-objective level rather than through the auxiliary regularizers
prior work relies on.

\item We replace static dispatch with adaptive retrieval at serving
time. Presence gating, followed by query-level NMS, activates
$\tilde{K} \le K$ queries per request while keeping performance on par
with all-$K$ retrieval.

\item Extensive offline experiments against learned multi-interest
baselines, and an online A/B test in Snap DPA, demonstrate the
effectiveness of \method{} in a production recommender.
\end{itemize}

\section{Related Work}
\label{sec:related}

\noindent\textbf{Multi-interest retrieval.}
Multi-interest recommenders represent each user with multiple
embeddings to capture the diverse interests reflected in their
behavior history. Existing methods differ mainly in how these
embeddings are constructed. MIND~\cite{li2019mind} uses capsule
routing~\cite{sabour2017capsules}; ComiRec uses multi-head
self-attention~\cite{cen2020comirec}; SINE activates a sparse subset
of global prototypes~\cite{tan2021sine}; MVKE introduces
virtual-kernel experts~\cite{xu2022mvke}; DCM performs differentiable
clustering~\cite{fan2025dcm}; KuaiFormer appends learnable query
tokens to a transformer over the behavior
sequence~\cite{liu2024kuaiformer}; and PinnerSage derives interests
from offline clustering of user behaviors~\cite{pal2020pinnersage}.
Subsequent work further improves these models through hard-negative
mining and contrastive learning
objectives~\cite{zhang2022re4,xie2023remi}. We refer readers to Li et al.~\cite{li2025misurvey} for a broader survey of multi-interest recommendation.

Despite these advances, two challenges remain in production
retrieval. First, many methods use argmax-based target-to-query
assignment~\cite{li2019mind,cen2020comirec}. Multiple targets may then
be assigned to the same query, while unmatched queries receive no
training signal, leading to \textbf{interest collapse}. Second, most
multi-interest retrievers use \textbf{static dispatch}, issuing a
fixed number of ANN queries for every request regardless of how many
user interests are active. Both persist in the deployed systems
closest to \method{}. KuaiFormer~\cite{liu2024kuaiformer} at Kuaishou
is the nearest in architecture, a transformer over the behavior
sequence with learnable query tokens, and shows that this design is
practical at production scale; \method{} adopts the same family but
replaces its argmax assignment and fixed ANN budget.
DCM~\cite{fan2025dcm} reduces interest collapse through
single-assignment clustering but still relies on static dispatch.
\method{} addresses both challenges by formulating multi-interest
retrieval as set prediction. Hungarian matching provides
one-to-one target assignment during training, while a per-query
presence head provides explicit no-interest supervision and
dynamically gates ANN queries at serving time.

\noindent\textbf{Set prediction, multi-vector retrieval, and selection.}
Set prediction produces a variable-size output set from a fixed bank
of learnable queries. DETR~\cite{carion2020detr} popularized this
formulation for object detection; Set
Transformer~\cite{lee2019set}, Perceiver
IO~\cite{jaegle2022perceiverio}, and Q-Former~\cite{li2023blip2}
extend the learnable-query paradigm to other domains. \method{}
adapts this paradigm to retrieval-stage recommendation. Specifically, \method{} inherits DETR's learnable query bank, Hungarian assignment, and explicit \emph{absence} target for unmatched queries, but applies them to a different input and a different prediction target. DETR predicts boxes in a bounded continuous space, supervised by
exhaustive annotation. Here the prediction is a vector in a frozen
item-embedding space, supervised contrastively, and the ground truth is a
user's engagement in a future window, so an unmatched query indicates only
that no further interest appears in that window. The presence score also
acts at a different point: in detection every query is decoded and the
absence class filters the output, whereas here it determines how many of
the $K$ ANN searches are issued.

Unlike ColBERT-style multi-vector
retrieval~\cite{khattab2020colbert,santhanam2022colbertv2}, where
each \emph{document} is represented by many vectors with late
interaction at scoring time, \method{} keeps every \emph{item} as a
single vector and places multiplicity only on the \emph{user} side,
so existing single-vector ANN
indices~\cite{guo2020scann,johnson2021faiss} are reused without
modification. Compared to differentiable top-$M$ selection (e.g.,
Gumbel-Softmax~\cite{jang2017gumbel}), the presence head supports a
variable number of active queries per request and avoids the
temperature schedule and straight-through estimator that top-$M$
sampling requires.

\section{Method}
\label{sec:method}

\begin{figure*}[t]
  \centering
  \includegraphics[width=\linewidth]{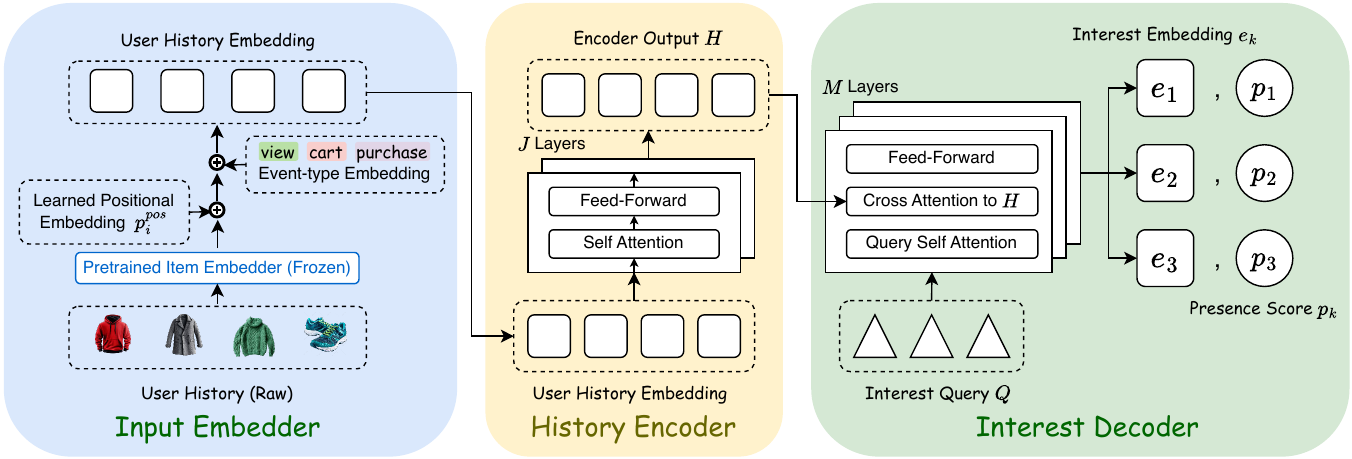}
  \caption{\method{} pipeline. A transformer encoder reads the user
    history, a $K$-query transformer decoder emits per-query retrieval
    embeddings and presence scores, and training uses Hungarian
    matching to align engaged targets with queries. At inference,
    presence gating and query-level NMS select the active ANN queries,
    whose results are merged by maximum score.}
  \Description{Method framework diagram: user-history items pass through
    a frozen content-embedding lookup and a transformer encoder, K learnable
    interest queries pass through a transformer decoder that cross-attends
    to the encoded history, and each query emits a retrieval embedding and
    a presence score; Hungarian matching aligns queries with engaged-target
    items at training time.}
  \label{fig:architecture}
\end{figure*}

\subsection{Problem statement}
\label{sec:problem}

We frame multi-interest retrieval as set prediction over $K$
learnable queries. Given a user's chronological engagement history
$\mathcal{H}_u = [v_1, v_2, \ldots, v_L]$, where each $v_i$ pairs an
engaged product with an engagement type, the task is to recover the
user's multiple interests. The most recent high-intent products
(e.g., purchase, add-to-cart, swipe-up) form the target set
$\mathcal{T}_u = \{t_1, \ldots, t_W\} \subset \mathcal{V}$, where
$\mathcal{V}$ is the item universe. Earlier engaged products serve
as model inputs. The number of targets $W$ varies across users and
is unknown at inference time.

The retrieval task is to return a top-$N$ candidate list
$\mathcal{C}_u \subseteq \mathcal{V}$ with high recall against
$\mathcal{T}_u$ under a serving budget: at most $K$ interest
embeddings may query the item index per request. \method{} therefore
predicts a set of interest representations
$\mathcal{P}_u = \{(e_k, p_k)\}_{k=1}^{K}$, where
$e_k \in \mathbb{R}^d$ is an interest embedding and $p_k \in [0, 1]$
is a presence score indicating whether query $k$ corresponds
to a current user interest.

\subsection{Architecture}
\label{sec:architecture}
Figure~\ref{fig:architecture} illustrates the overall framework: an
input embedder, a history encoder, and an interest decoder with two
per-query output heads, detailed in the following paragraphs.

\noindent\textbf{Input embedder.}
Each history product $v_i$ is embedded by a frozen pretrained
content-embedding model from its metadata (title, brand, category,
image) into a content embedding $\bar{x}_i \in \mathbb{R}^d$. We add a learned positional
embedding $p_i^{\text{pos}}$ and an
\emph{$\alpha$-gated event-type
embedding} $\alpha\,c(\text{type}_i) \in \mathbb{R}^d$ to form the
per-item input. The
event-type embeddings are learned from scratch, and the scalar
$\alpha$ is initialized at $0$ and learned directly, so the model
starts from the pure content-embedding space and gradually
incorporates engagement signals:
\begin{equation}
  x_i = \bar{x}_i + p_i^{\text{pos}} + \alpha\,c(\text{type}_i)
  \in \mathbb{R}^d.
\end{equation}

\noindent\textbf{History encoder.}
A standard $J$-layer transformer encoder
$\mathcal{E}_\theta$~\cite{vaswani2017attention} processes the
history of per-item inputs $[x_1, \ldots, x_L]$ into contextualized
representations:
\begin{equation}
  H = \mathcal{E}_\theta\bigl([x_1, \ldots, x_L]\bigr)
  \in \mathbb{R}^{L \times d}.
  \label{eq:encoder}
\end{equation}

\noindent\textbf{Interest decoder with $K$ learnable queries.}
\method{} maintains $K$ learnable interest queries, shared across
users, $Q^{(0)} = [q_1,\ldots,q_K] \in \mathbb{R}^{K \times d}$,
randomly initialized and trained jointly with the rest of the model. An
$M$-layer transformer decoder $\mathcal{D}_\phi$ alternates between
self-attention among queries and cross-attention to the encoded
history. Omitting residual connections and layer normalization for
compactness, one decoder layer is
\begin{align}
  \tilde Q^{(\ell)} &= \mathrm{SelfAttn}\bigl(Q^{(\ell-1)}\bigr), \\
  \bar Q^{(\ell)} &= \mathrm{CrossAttn}\bigl(\tilde Q^{(\ell)}, H, H\bigr), \\
  Q^{(\ell)} &= \mathrm{FFN}\bigl(\bar Q^{(\ell)}\bigr),
\end{align}
for $\ell = 1,\ldots,M$. Query self-attention lets interest queries
exchange information before reading the history, which encourages
different queries to specialize rather than independently chase the
same dominant signal. The decoder output
$Z = Q^{(M)} \in \mathbb{R}^{K \times d}$ is passed to two per-query
heads.

\noindent\textbf{Two output heads per query.}
For each decoder-output row $z_k$ of $Z = Q^{(M)}$, \method{}
predicts a normalized retrieval embedding $e_k$ and a presence
score $p_k$:
\begin{align}
  e_k &= \frac{W_e\,\mathrm{LN}(z_k) + b_e}
              {\bigl\|W_e\,\mathrm{LN}(z_k) + b_e\bigr\|_2}
        \in \mathbb{R}^d, \\
  p_k &= \sigma\bigl(w_p^\top \mathrm{LN}(z_k) + b_p\bigr)
        \in [0,1],
\end{align}
where $\sigma(\cdot)$ is the logistic sigmoid. The $L_2$
normalization makes training and serving consistent. The scalar
presence head estimates whether the query is active for this
request.

\subsection{Set-prediction training}
\label{sec:training}

For a training example, target product identifiers are deduplicated
by product id (pid). If the same pid appears with multiple event
types, we keep the strongest-intent event type (priority order:
purchase $>$ add-to-cart $>$ swipe-up). The target products are then
capped at $W_{\max}=15$ by strongest-intent priority. The model
produces $K$ interest embeddings $\{e_k\}_{k=1}^{K}$, and the same
frozen pretrained model from §\ref{sec:architecture} encodes each
target product $t_w$ into a target embedding $\bar{t}_w \in
\mathbb{R}^d$, giving the target set $\{\bar{t}_w\}_{w=1}^{W}$ where
$W \le W_{\max}$ and may exceed $K$.

\noindent\textbf{Hungarian matching.}
We compute a one-to-one
assignment between predicted queries and target items. Let
$\mathcal{A}_{K,W}$ be the set of matchings
$A \subseteq [K]\times[W]$ with $|A|=\min(K,W)$ and no repeated query
or target. We solve
\begin{equation}
  \begin{aligned}
  \hat{A} &= \argmin_{A \in \mathcal{A}_{K,W}}
              \sum_{(k,w)\in A}\,\Gamma_{k,w}, \\
  \Gamma_{k, w}
           &= -\,\lambda_{\text{emb}}\,e_k^\top \bar{t}_w
              -\,\lambda_{\text{cls}}\,\log\sigma\!\bigl(\ell_k\bigr),
  \end{aligned}
  \label{eq:hungarian-cost}
\end{equation}
where $\ell_k$ is the pre-sigmoid presence logit. The embedding term
prefers geometrically close query-target pairs; the class term biases
the matcher toward queries the model already predicts as present. We
use $\lambda_{\text{emb}} = 1$ and $\lambda_{\text{cls}} = 0.5$, with
the class term linearly warmed up over the first $10\%$ of training
steps so that early matching is driven primarily by embedding
geometry. For $K=10$, each user's assignment is a small
$K \times W$ problem and the solve itself is negligible.
Let
$\mathcal{M}_u = \{k : (k,w) \in \hat{A}\}$ denote the matched query
indices, with $|\mathcal{M}_u| = \min(W,K)$, and let
$\mathcal{U}_u = [K] \setminus \mathcal{M}_u$ denote unmatched
(\enquote{no-interest}) queries.

\noindent\textbf{InfoNCE retrieval loss.}
For each matched pair $(k,w) \in \hat{A}$, we apply an in-batch
sampled-softmax (InfoNCE~\cite{oord2018cpc}) loss. The assigned target $\bar{t}_w$ is the
positive, and all valid target embeddings gathered across the
data-parallel mini-batch form the contrastive pool $\mathcal{N}$, which
therefore contains $\bar{t}_w$ itself:
\begin{equation}
  \mathcal{L}_{\text{InfoNCE}}(u) = -\frac{1}{|\hat{A}|}
  \sum_{(k,w)\in\hat{A}}
  \log\frac{\exp\bigl(e_k^\top \bar{t}_w / \tau_r\bigr)}
           {\sum_{\bar{t}^\prime \in \mathcal{N}}
            \exp\bigl(e_k^\top \bar{t}^\prime / \tau_r\bigr)},
  \label{eq:infonce}
\end{equation}
where $\tau_r$ is the InfoNCE temperature and padded target entries
are masked out. Only matched queries participate in
$\mathcal{L}_{\text{InfoNCE}}$; unmatched queries are not pulled
toward any target direction.

\noindent\textbf{Presence loss.}
The matcher identifies which queries are used on the example, and the
presence head learns to predict that assignment. Given presence score $p_k$ and binary label
$y_k = \mathbf{1}[k \in \mathcal{M}_u]$,
\begin{equation}
  \mathcal{L}_{\text{Pres}}(u) = -\frac{1}{K}\sum_{k=1}^{K}
    \bigl[ y_k \log p_k + (1 - y_k)\log(1 - p_k) \bigr].
  \label{eq:presence}
\end{equation}
The retrieval and presence losses provide complementary supervision.
Matched queries are pulled toward their assigned targets, whereas
unmatched queries receive only a $y_k{=}0$ presence target. This
absence signal is the mechanism missing from argmax-trained
multi-interest models: when targets independently choose their
closest query, unmatched queries receive no direct supervision and can
collapse or drift.

\noindent\textbf{Margin diversity loss.}
We add a light pairwise repulsion term to discourage active queries
from becoming nearly identical. Let
$a_k=\mathbf{1}[p_k>0.5]$ be a detached hard activity mask; gradients
do not flow through the threshold. We define
\begin{equation}
  \mathcal{L}_{\text{Div}}(u) =
  \frac{\sum_{i \neq j} a_i\,a_j\,\max\!\bigl(0,\, e_i^\top e_j - m\bigr)}
       {\max\!\bigl(1,\, \sum_{i \neq j} a_i\,a_j\bigr)},
  \label{eq:diversity}
\end{equation}
with margin $m = 0.3$. The loss is zero when fewer than two queries
are active. Otherwise, it penalizes only active pairs whose cosine
similarity exceeds $m$, leaving already separated queries unchanged.
Unlike ComiRec's controllability term~\cite{cen2020comirec}, which is
applied during inference, this term shapes the learned retrieval
geometry during training.

\noindent\textbf{Total loss.}
For a mini-batch of users $\mathcal{B}$, the total training loss is
\begin{equation}
  \mathcal{L} = \frac{1}{|\mathcal{B}|}\sum_{u \in \mathcal{B}}
    \bigl[\,\lambda_R\, \mathcal{L}_{\text{InfoNCE}}(u)
        + \lambda_P\, \mathcal{L}_{\text{Pres}}(u)
        + \lambda_D\, \mathcal{L}_{\text{Div}}(u)\,\bigr],
  \label{eq:total-loss}
\end{equation}
with loss weights $\lambda_R, \lambda_P, \lambda_D$ specified in
§\ref{sec:setup}.

\subsection{Inference}
\label{sec:inference}

\begin{figure}[t]
  \centering
  \includegraphics[width=\linewidth]{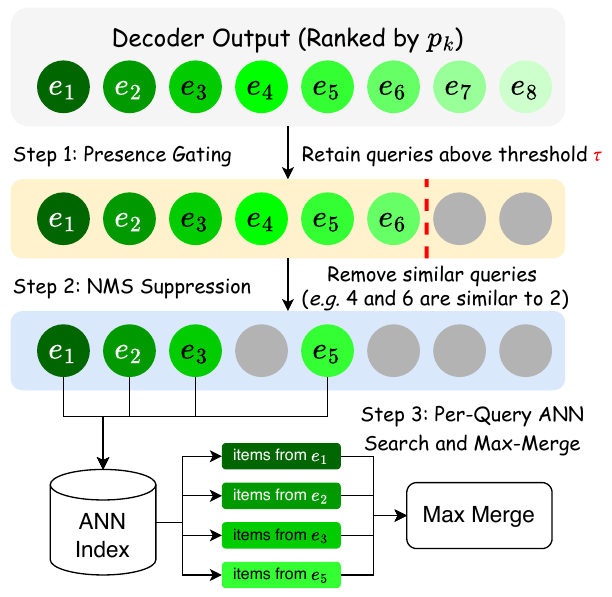}
  \caption{Inference-time gating and per-query ANN. Presence gating at
    threshold $\tau$ retains active queries, cosine-NMS at
    $\delta_{\text{NMS}}$ removes redundant survivors, each survivor
    issues one ANN search, and returned items are merged by max-score.}
  \Description{Flow diagram showing K queries entering a presence-gating
    step that drops low-presence queries, an NMS step that drops
    redundant queries, per-query ANN search, and a max-merge step
    producing the final candidate list.}
  \label{fig:inference}
\end{figure}

At serving time, \method{} performs a single forward pass that
encodes the user history and decodes the $K$ interest queries,
yielding $\{(e_k,p_k)\}_{k=1}^{K}$. It then converts the fixed query
bank into a dynamic set of $\tilde{K} \le K$ ANN queries through
three lightweight steps that together replace static dispatch with
dynamic $\tilde{K}$ retrieval (see
Figure~\ref{fig:inference}).

\noindent\textbf{Step 1: Presence gating.}
We first retain queries whose presence exceeds a threshold $\tau$:
\begin{equation}
  \mathcal{S}_1 = \{\, k \in [K]\;:\; p_k > \tau \,\}.
\end{equation}
If no query passes the threshold, we keep the query with the largest
presence score as a fallback.

\noindent\textbf{Step 2: Query-level non-maximum suppression.}
We sort $\mathcal{S}_1$ by descending $p_k$ and apply NMS over
query-query cosine similarity. A query is removed if its cosine
similarity to any higher-presence retained query exceeds
$\delta_{\text{NMS}}$; we use $\delta_{\text{NMS}}=0.9$. The surviving
set is $\mathcal{S}_2 \subseteq \mathcal{S}_1$, with
$|\mathcal{S}_2|=\tilde{K}$.

\noindent\textbf{Step 3: Per-query ANN search and max-merge.}
Each surviving query $k \in \mathcal{S}_2$ retrieves its top-$M_q$
items from the item index. We set
$M_q = \lceil N / \tilde{K} \rceil$ so that the total retrieval depth
remains approximately constant as $\tilde{K}$ changes. Items returned
by multiple queries are merged by maximum score:
\begin{equation}
  s(t \mid u) = \max_{k \in \mathcal{S}_2} e_k^\top \bar{t},
\end{equation}
and the top-$N$ items by $s(t \mid u)$ form the candidate list.

\section{Experiments}
\label{sec:experiments}

\subsection{Experimental Setup}
\label{sec:setup}

\noindent\textbf{Dataset.}
We evaluate \method{} on user-product engagement logs from Snap DPA.
The advertiser catalog contains hundreds of millions of products,
and the event stream includes views, swipe-ups, add-to-carts, and
purchases. For each user, we form the target set $\mathcal{T}_u$
from all high-intent events (such as purchase, add-to-cart, and
swipe-up) that fall within the last 3-day window. Targets are
deduplicated by product id and capped at $W_{\max}{=}15$ items, which retains all high-intent targets for most users in the training data.
The history sequence is every event whose timestamp lies strictly before
the start of the window, truncated to the most recent $L{=}40$
events. Users are partitioned into disjoint train, validation, and
test sets. Item embeddings are
computed offline by a separately pretrained content-embedding model,
then frozen and stored as $L_2$-normalized vectors throughout
\method{} training and evaluation.

\begin{table*}[t]
  \caption{Comparison of \method{} against four \emph{learned}
  multi-interest retrieval methods. All rows share the same training
  data, the same frozen item embeddings, $K{=}10$, the same optimizer
  and batch size, and the same per-request retrieval budget; they
  differ only in the interest-extraction architecture and the
  target-to-query assignment. Baselines have no presence concept, so
  they dispatch all $K$ queries ($\tilde{K}{=}K$), which is the static
  dispatch they practice in production. Every method is trained until it
  converges, each to its own point of convergence rather than to a shared
  step count, since the model classes differ substantially in how long they keep improving.
  The \method{}~(argmax) row replaces Hungarian matching with
  argmax assignment and is analyzed in \S\ref{sec:ablation}. Each
  metric is normalized so that \method{} equals $100\%$.}
  \label{tab:learned-baselines}
  \small
  \begin{tabular}{lrrrrrrrr}
    \toprule
    Method & $\tilde{K}$ & R@1 & R@5 & R@10 & R@100 & HR@100 & NDCG@100 & MRR \\
    \midrule
    MIND~\cite{li2019mind}              & $10.00$ & $45.7\%$  & $50.7\%$  & $52.9\%$  & $61.7\%$  & $58.2\%$  & $54.3\%$  & $44.4\%$  \\
    ComiRec-SA~\cite{cen2020comirec}    & $10.00$ & $67.4\%$  & $83.5\%$  & $89.8\%$  & $99.9\%$  & $99.6\%$  & $89.8\%$  & $80.0\%$  \\
    DCM~\cite{fan2025dcm}               & $10.00$ & $26.0\%$  & $34.5\%$  & $37.4\%$  & $49.3\%$  & $48.8\%$  & $39.3\%$  & $29.1\%$  \\
    KuaiFormer~\cite{liu2024kuaiformer} & $10.00$ & $43.0\%$  & $47.6\%$  & $52.0\%$  & $62.8\%$  & $64.1\%$  & $52.3\%$  & $42.2\%$  \\
    \midrule
    \method{} (argmax)                  & $2.16$  & $62.1\%$  & $63.2\%$  & $63.5\%$  & $69.1\%$  & $73.2\%$  & $63.1\%$  & $56.4\%$  \\
    \method{}                           & $6.70$  & $100\%$   & $100\%$   & $100\%$   & $100\%$   & $100\%$   & $100\%$   & $100\%$   \\
    \bottomrule
  \end{tabular}
\end{table*}

\noindent\textbf{Evaluation pool.}
All offline results are reported on a held-out \emph{test} set of around
$1$M users, yielding a target pool of around $5.3$M products. The
validation split also includes around $1$M users and is used to choose the presence threshold $\tau$, the NMS threshold $\delta_{\text{NMS}}$, and each method's converged checkpoint, so no reported number informs those choices.

\noindent\textbf{Metrics.}
We report $\text{R}@N$ (recall at top-$N$ retrieved products) for
$N\in\{1, 5, 10, 100\}$, $\text{HR}@100$ (hit rate),
$\text{NDCG}@100$, and $\text{MRR}$ (mean reciprocal rank).
For serving analysis we additionally report $\tilde{K}$, the average
number of ANN queries per request after presence gating and NMS. Per
Snap's policy, we report only relative metric lifts.

\begin{table*}[t]
  \caption{Component ablation. Each row removes exactly one mechanism;
  architecture, data and the
  remaining losses are held fixed, on the same converged schedule as
  Table~\ref{tab:learned-baselines}. Rows are normalized to \method{} and
  $\tilde{K}$ is absolute at $\tau{=}0.3$. Each mechanism is discussed in
  \S\ref{sec:ablation}.}
  \label{tab:ablation}
  \small
  \begin{tabular}{lrrrrrrrr}
    \toprule
    Variant & $\tilde{K}$ & R@1 & R@5 & R@10 & R@100 & HR@100 & NDCG@100 & MRR \\
    \midrule
    \method{} (full)             & $6.70$ & $100\%$   & $100\%$   & $100\%$   & $100\%$   & $100\%$   & $100\%$   & $100\%$   \\
    \quad $-$ presence loss      & $8.23$ & $97.1\%$ & $90.4\%$ & $91.4\%$ & $91.3\%$ & $88.4\%$ & $92.9\%$ & $97.4\%$ \\
    \quad $-$ diversity loss     & $7.55$ & $88.5\%$ & $90.6\%$ & $93.2\%$ & $91.8\%$ & $91.5\%$ & $93.3\%$ & $95.7\%$ \\
    \quad argmax assignment      & $2.16$ & $62.1\%$  & $63.2\%$  & $63.5\%$  & $69.1\%$  & $73.2\%$  & $63.1\%$  & $56.4\%$  \\
    \bottomrule
  \end{tabular}
\end{table*}

\begin{table*}[t]
  \caption{K sweep experiment. All rows share architecture,
  optimization, and training duration; only $K$ varies. Each metric
  is normalized so that the single-query ($K{=}1$) configuration
  equals $100\%$. The second column reports $\tilde{K}$, the
  average number of presence-active queries.}
  \label{tab:main-results}
  \small
  \begin{tabular}{lrrrrrrrr}
    \toprule
    $K$ & $\tilde{K}$ & R@1 & R@5 & R@10 & R@100 & HR@100 & NDCG@100 & MRR \\
    \midrule
    $1$  & $1.00$ & $100\%$   & $100\%$   & $100\%$   & $100\%$   & $100\%$   & $100\%$   & $100\%$   \\
    $3$  & $3.00$ & $108.4\%$ & $133.5\%$ & $137.5\%$ & $137.2\%$ & $110.8\%$ & $130.7\%$ & $112.4\%$ \\
    $5$  & $4.72$ & $113.2\%$ & $145.8\%$ & $151.1\%$ & $148.6\%$ & $113.3\%$ & $141.6\%$ & $117.2\%$ \\
    $7$  & $5.75$ & $112.3\%$ & $147.6\%$ & $154.7\%$ & $151.9\%$ & $114.2\%$ & $144.2\%$ & $117.8\%$ \\
    $10$ & $6.70$ & $113.0\%$ & $148.3\%$ & $155.0\%$ & $152.8\%$ & $114.1\%$ & $145.8\%$ & $118.9\%$ \\
    $15$ & $7.59$ & $112.8\%$ & $148.6\%$ & $155.1\%$ & $152.6\%$ & $114.1\%$ & $145.9\%$ & $118.4\%$ \\
    \bottomrule
  \end{tabular}
\end{table*}

\noindent\textbf{Implementation.}
The history encoder is a $2$-layer transformer with $8$ attention heads,
feed-forward width $d_{\text{ff}}=1024$, and dropout $0.1$; the interest
decoder has $6$ layers with the same width, heads, and dropout, with hidden size $d=128$ and approximately $3$M learnable parameters.

We train with AdamW~\cite{loshchilov2019adamw} (peak learning rate $5\times 10^{-4}$, $4000$
warm-up steps, weight decay $0.01$, cosine decay) on $8$ A100-40GB
GPUs with batch size $1536$ users per GPU and bfloat16 mixed
precision. The in-batch InfoNCE pool gathers target embeddings
across GPUs and masks padded target entries. The InfoNCE temperature is
$\tau_r=0.05$, the diversity margin is $m=0.3$, and the loss weights
are $\lambda_R=1.0$, $\lambda_P=0.5$, $\lambda_D=0.1$, with the
Hungarian class term $\lambda_{\text{cls}}=0.5$ linearly warmed up
over the first $10\%$ of steps. Unless otherwise stated, inference
uses presence threshold $\tau=0.3$, NMS threshold
$\delta_{\text{NMS}}=0.9$, and $K{=}10$ queries.

\begin{table*}[t]
  \caption{Ablation of the $\alpha$-gated event-type embedding:
  percentage change from a no-event-type-embedding baseline at three
  training durations. All other architecture and training settings
  are held fixed.}
  \label{tab:event-emb}
  \small
  \begin{tabular}{lrrrrrrr}
    \toprule
    Steps & $\Delta$R@1 & $\Delta$R@5 & $\Delta$R@10 & $\Delta$R@100 & $\Delta$HR@100 & $\Delta$NDCG@100 & $\Delta$MRR \\
    \midrule
    $30$K  & $+5.1\%$ & $+2.6\%$ & $+3.0\%$ & $+1.7\%$ & $+0.5\%$ & $+2.9\%$ & $+3.0\%$ \\
    $60$K  & $+7.6\%$ & $+5.6\%$ & $+4.0\%$ & $+1.8\%$ & $+0.5\%$ & $+4.3\%$ & $+5.3\%$ \\
    $120$K & $+8.2\%$ & $+6.9\%$ & $+5.0\%$ & $+3.2\%$ & $+1.6\%$ & $+5.2\%$ & $+5.6\%$ \\
    \bottomrule
  \end{tabular}
\end{table*}

\subsection{Offline Experiments}
\label{sec:offline}

\subsubsection{Comparison with learned multi-interest baselines.}
We compare \method{} against four learned multi-interest retrievers
spanning the designs surveyed in \S\ref{sec:related}:
MIND~\cite{li2019mind}, ComiRec-SA~\cite{cen2020comirec},
DCM~\cite{fan2025dcm}, and KuaiFormer~\cite{liu2024kuaiformer} on the Snap DPA evaluation pool.
Because \method{} consumes a \emph{frozen} pretrained item tower whereas
all four learn item embeddings jointly with the user side, this is a controlled
comparison of interest-extraction mechanisms rather than an end-to-end
reproduction of published results.
For the experiment comparison, ComiRec-SA and MIND are ported
from the authors' reference implementation\footnote{\url{https://github.com/THUDM/ComiRec}}; DCM and
KuaiFormer have no public code and are implemented from their papers. All
four use the same output head as \method{}, and MIND's ReLU output layer
is dropped because a frozen item tower would otherwise confine queries to
the non-negative orthant. Each method is trained until it
converges on the validation set, so the training length is per method
rather than shared: MIND, ComiRec-SA and DCM peak within $60$K steps, KuaiFormer at roughly $100$K and \method{} at $160$K.

\method{} is strongest on every metric, and the margins are widest
at the top of the ranking, which is what the downstream ranker consumes:
the best baseline, ComiRec-SA, reaches $67.4\%$ of \method{}'s R@1 and
$80.0\%$ of its MRR, yet becomes competitive at depth ($99.9\%$ of
R@100), so attention-based routing recovers a similar candidate
\emph{pool} but orders it less well. \method{} does this while issuing
$6.70$ ANN queries per request against the baselines' $10$, a $33\%$
smaller budget.

MIND and DCM fall furthest behind, at $52.9\%$ and $37.4\%$ of
\method{}'s R@10. Both construct interests geometrically, by capsule
routing and by differentiable clustering respectively, and both were
designed to learn the item tower jointly with the user side; DCM
additionally had to be reimplemented from its paper. We therefore read
these rows as evidence about how those mechanisms behave under a frozen
item space, not as a verdict on the published systems.

KuaiFormer is worth isolating: it is the closest architecture to
\method{}, a transformer over the history with learnable query tokens,
yet reaches only $52.0\%$ of \method{}'s R@10. Both are deep transformers
of comparable capacity on identical data and item embeddings, so the gap
points to how the queries are supervised, which the \method{}~(argmax)
row isolates directly (\S\ref{sec:ablation}).

\subsubsection{Component ablation.}
\label{sec:ablation}
Table~\ref{tab:ablation} removes one mechanism at a time: the
target-to-query assignment, the presence loss, and the diversity loss.
Replacing Hungarian matching with per-target argmax costs $36.5\%$ of R@10 and
$43.6\%$ of MRR and drops the mean number of presence-active queries from
$6.70$ to $2.16$ of $10$: targets compete for whichever query is already
closest, the rest never receive a retrieval gradient, and the model
settles on roughly two effective interests, illustrating the impact of interest collapse. 

Both loss terms also cost recall when removed: R@10 falls to
$91.4\%$ without the presence loss and $93.2\%$ without diversity.
Each term does distinct work. The presence loss is what makes gating
learnable: left unsupervised, $\sigma(\ell_k)$ carries no information about
whether a query is warranted, so thresholding at $\tau{=}0.3$ admits
$8.23$ queries rather than $6.70$ and still loses recall, which is more
calls for less return. The diversity term acts on query redundancy, and
mean pairwise cosine among the $K$ query embeddings rises from $0.298$ to
$0.321$ when it is removed. Across the three, the assignment dominates: it
costs $36.5\%$ of R@10 against $8.6\%$ and $6.8\%$ for the presence and
diversity terms, so one-to-one matching carries the retrieval quality
while the two losses contribute smaller, separate effects.



\begin{figure*}[!t]
  \centering
  \includegraphics[width=0.77\linewidth]{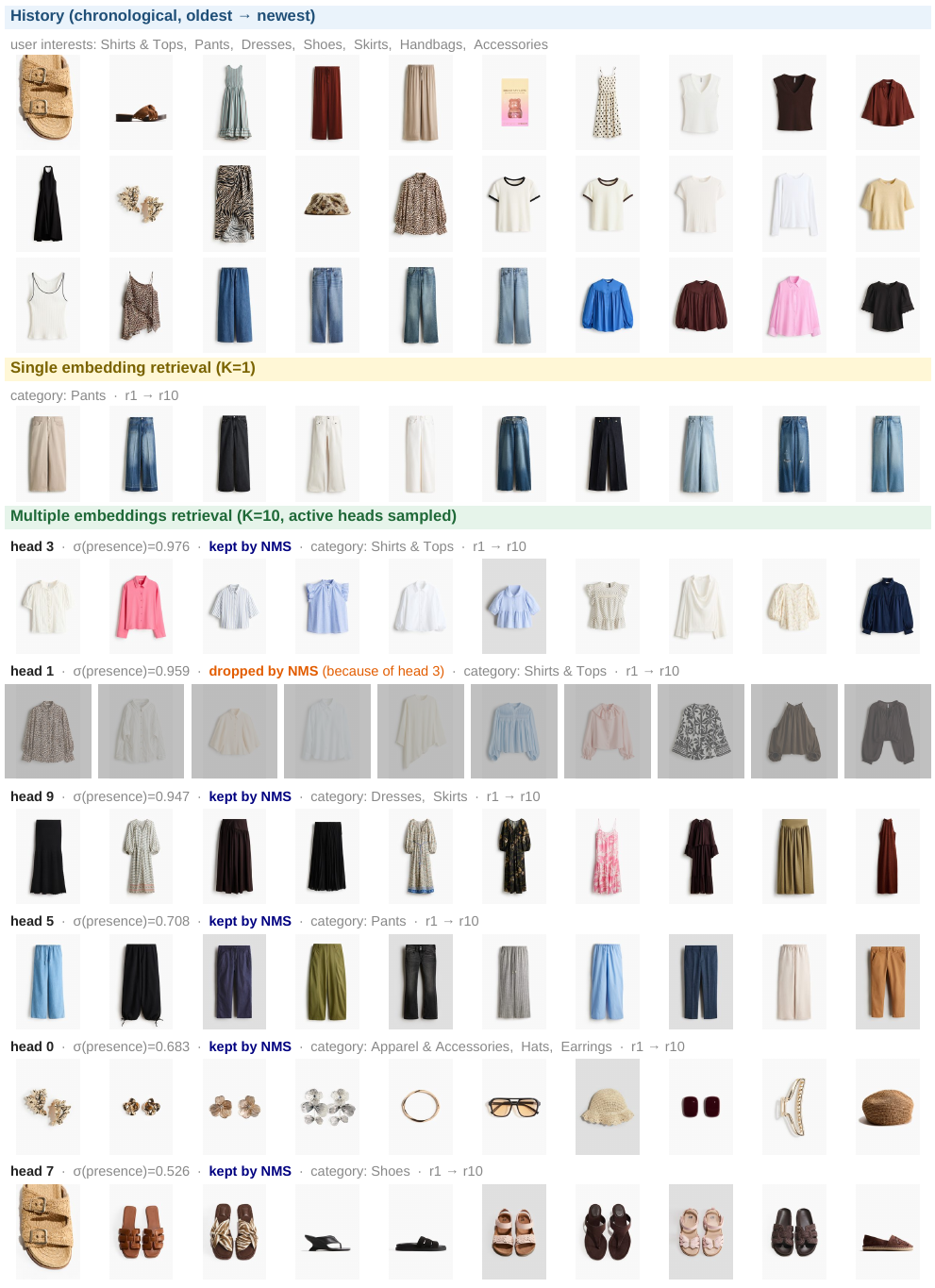}
  \caption{Qualitative retrieval comparison for one user. The figure
  labels \enquote{Single embedding ($K{=}1$)} and \enquote{Multiple
  embeddings ($K{=}10$)} correspond to the single-query baseline and
  \method{} respectively; per-query rows are labeled
  \enquote{head $k$}, each corresponding to one of \method{}'s $K$
  interest queries with its $\sigma(\text{presence})$ score from the
  gating step. The user's history spans shirts \& tops, pants,
  dresses, shoes, skirts, handbags, and accessories. The
  single-query baseline collapses onto pants; \method{}'s five active queries each
  specialize in a distinct category, while a near-duplicate query
  (head~1) is removed by NMS for overlapping with head~3.}
  \Description{Qualitative retrieval figure with three sections:
  history of user-engaged products spanning fashion categories,
  single-query baseline retrievals (all pants), and \method{}
  per-query retrievals across shirts, dresses, pants, accessories,
  and shoes, plus one head dropped by NMS for redundancy with the
  shirts head.}
  \label{fig:qualitative}
\end{figure*}

\begin{table*}[t]
  \caption{Training-data scaling sweep.
  All rows share architecture and training recipe, and each is trained
  to convergence; only the number of
  training users varies. Each metric is normalized so that the smallest scale ($100$K
  users) equals $100\%$.}
  \label{tab:data-scaling}
  \small
  \begin{tabular}{lrrrrrrr}
    \toprule
    Train users & R@1 & R@5 & R@10 & R@100 & HR@100 & NDCG@100 & MRR \\
    \midrule
    $100\text{K}$ & $100\%$   & $100\%$   & $100\%$   & $100\%$   & $100\%$   & $100\%$   & $100\%$   \\
    $1\text{M}$         & $179.1\%$ & $181.2\%$ & $179.2\%$ & $166.9\%$ & $145.1\%$ & $175.1\%$ & $171.0\%$ \\
    $10\text{M}$        & $247.5\%$ & $262.5\%$ & $258.4\%$ & $218.5\%$ & $169.7\%$ & $243.3\%$ & $231.2\%$ \\
    \bottomrule
  \end{tabular}
\end{table*}

\subsubsection{K sweep.} Having established that \method{} improves on prior learned
multi-interest retrievers, we now vary the number of decoder
queries $K$ while holding the rest of the configuration fixed
(Table~\ref{tab:main-results}). The $K{=}1$ row serves as the
single-query configuration.

Three patterns stand out. \emph{First, increasing $K$ produces large
recall gains.} $K{=}10$ reaches $155\%$ of the $K{=}1$ R@10 and
$153\%$ of the $K{=}1$ R@100. The largest jump is
from $K{=}1$ to $K{=}3$, indicating that even a small number of
additional interest queries captures most of the available gain.
\emph{Second, gains are larger at deeper retrieval depths.} The lift
from $K{=}1$ to $K{=}10$ is wider on R@10/R@100
($+55\%$/$+53\%$) than on R@1/HR@100 ($+13\%$/$+14\%$), consistent
with multi-interest retrieval broadening the candidate surface
rather than only improving the top hit.
\emph{Third, gains gradually saturate as $K$ increases.} $K{=}7$
already reaches $99.8\%$ of $K{=}10$'s R@10, and $K{=}15$ is within $\pm 0.5\%$ of
$K{=}10$ across all columns. The $\tilde{K}$ column makes this
explicit: $\tilde{K}$ grows sublinearly with $K$ ($4.72$ at $K{=}5$,
$6.70$ at $K{=}10$, only $7.59$ at $K{=}15$), so even when $15$
queries are available the model commits to fewer than $8$ on
average; the
effective number of interests is bounded by the data, not by $K$.

\subsubsection{Event-type embedding.}
\label{sec:event-emb-abl}
Table~\ref{tab:event-emb} isolates the $\alpha$-gated event-type
embedding of \S\ref{sec:architecture}.
The event signal helps consistently, and its effect grows with training
duration. Because $\alpha$ is initialized at zero, the model first
matches the no-event baseline and then gradually learns how much
event-type information to add: the R@10 gain rises from $+3.0\%$ to
$+5.0\%$ between the shortest and longest schedule. The gain is
strongest near the top of the ranking ($\Delta$R@1$=+8.2\%$ vs.\
$\Delta$R@100$=+3.2\%$), which complements the $K$ sweep, where the
multi-interest gain is more visible at depth.

\subsubsection{Data scaling.}
Table~\ref{tab:data-scaling} sweeps the number of training users. Each metric is normalized so that the smallest scale ($100$K users) equals $100\%$. Performance scales smoothly with training data: $1$M users lifts the R@100 to $1.66\times$ and HR@100 to $1.45\times$, while $10$M users further improves the R@100 and HR@100 to $2.18\times$ and $1.69\times$, respectively.

\begin{table}[t]
  \caption{Serving ablation. One checkpoint served under three dispatch
  policies; no retraining. \emph{ANN calls} is the mean number of
  per-request ANN queries actually issued, and metrics are normalized so
  that static all-$K$ dispatch equals $100\%$. Presence gating removes
  $23\%$ of the ANN calls at $99.4\%$ of R@10 while improving R@1, and
  NMS at $\delta_{\text{NMS}}{=}0.90$ removes a further $1.03$ calls at
  $99.2\%$ of R@10, so the two together issue $6.70$ of $10$.}
  \label{tab:serving-ablation}
  \small
  \begin{tabular}{lrrrr}
    \toprule
    Dispatch policy & ANN calls & R@1 & R@10 & MRR \\
    \midrule
    static all-$K$ ($\tau{=}0$)             & $10.00$ & $100\%$   & $100\%$  & $100\%$   \\
    \; + presence gating ($\tau{=}0.3$)     & $7.73$  & $101.9\%$ & $99.4\%$ & $100.7\%$ \\
    \; \; + query-level NMS ($\delta{=}0.9$)   & $6.70$  & $101.8\%$ & $99.2\%$ & $100.7\%$ \\
    \bottomrule
  \end{tabular}
\end{table}

\subsubsection{Inference-time gating and dispatch reduction.}
\label{sec:serving-ablation}
Table~\ref{tab:serving-ablation} tests whether gating and NMS deliver the
dynamic $\tilde{K} \le K$ dispatch that replaces static all-$K$
retrieval, serving one $K{=}10$ checkpoint under three policies.
Gating accounts for most of the reduction ($10 \to 7.73$ calls of
the $10 \to 6.70$ total), and it slightly \emph{improves} R@1 and MRR,
because suppressing low-confidence queries
removes candidates that would otherwise dilute the merged list. The two
knobs are complementary rather than interchangeable: gating asks whether
the user has the interest at all, which the head is trained to answer,
whereas NMS applies a purely geometric test to whichever queries survive,
so each removes calls the other keeps. Loosening the gate below
$\tau{=}0.30$ buys little ($99.4\%\to100\%$ R@10 for $7.73\to10$ calls),
so we deploy $\tau{=}0.30$ with $\delta_{\text{NMS}}{=}0.90$.

\subsection{Online Experiment}
\label{sec:online}

We deployed \method{} as a new retrieval source in the
Snap DPA production stack and ran a live A/B test with the downstream
ranker held fixed. \method{} uses its default inference policy and its own ANN index built
from the same pretrained content-embedding model. The experiment ran on a
representative live-traffic slice for about one week; exact traffic
fractions, sample sizes, and significance thresholds are omitted per
Snap's policy. We report two analyses from this single A/B:
\textbf{(a)} the group-level lift of treatment over
control, capturing \method{}'s overall contribution to the production
stack; and \textbf{(b)} a source-level comparison of \method{} versus
the embedding-based item-to-item (I2I) matching retrieval source running alongside it in the treatment arm.

\noindent\textbf{(a) Overall contribution to the production stack.}
The control retains the current Snap DPA retrieval mix; the
treatment adds \method{} as a new retrieval source while keeping the
global retrieval quota unchanged. Adding \method{} lifts user
impressions by $+0.10\%$, CTR by $+0.21\%$, and CVR by $+3.11\%$
(Table~\ref{tab:online-ab}, column (a)).

\begin{table}[t]
  \caption{Online A/B results from a single live experiment.
  \textbf{(a)} Group-level engagement lifts of treatment
  (production stack + \method{}) over control (production stack).
  \textbf{(b)} Source-level engagement attribution within the
  treatment arm, comparing \method{} against the item-to-item (I2I)
  retrieval source running alongside it under matched per-source
  retrieval quota. Downstream ranker held fixed across
  both analyses.}
  \label{tab:online-ab}
  \small
  \begin{tabular}{lrr}
    \toprule
    Metric & (a) $\Delta$ vs control & (b) $\Delta$ \method{} vs I2I \\
    \midrule
    CTR    & $+0.21\%$ & $+44\%$ \\
    CVR    & $+3.11\%$ & $+51\%$ \\
    \bottomrule
  \end{tabular}
\end{table}

\noindent\textbf{(b) Per-source comparison vs item-to-item retrieval.}
Item-to-item (I2I) retrieval dispatches a user's recent engaged
high-intent products directly as ANN seeds, sharing \method{}'s input
signal, frozen item embeddings, ANN index, and per-source retrieval
quota; the two differ only in how the query vectors are produced. Since
I2I already dispatches several seeds per request, this is a comparison
between two ways of choosing a small set of query vectors, and serving
cost is comparable because both issue a few single-vector ANN lookups per
request. Attributing engagement to each source within the treatment arm,
\method{} achieves $+44\%$ CTR and $+51\%$ CVR over I2I
(Table~\ref{tab:online-ab}, column (b)), isolating the value of the
learned encoder--decoder, Hungarian matching, and presence head over the
simpler heuristic of dispatching engaged history items directly.

\subsection{Qualitative Analysis}
\label{sec:qualitative}

Figure~\ref{fig:qualitative} visualizes the interest
collapse failure mode on one user: although the user's history
spans shirts \& tops, pants, dresses, shoes, skirts, handbags, and
accessories, the single-query ($K{=}1$) baseline returns only pants.
\method{}'s five active queries each specialize in a distinct
interest cluster, so the merged retrieval covers most of the user's interest mix.

\section{Conclusion}
\label{sec:conclusion}

We presented \method{} for multi-interest retrieval. The model uses
$K$ learnable interest queries, Hungarian matching against engaged
targets, and a presence head supervised by the matcher. Together,
these components address interest collapse at training time and
static dispatch at serving time: presence gating and query-level NMS
turn the fixed query bank into a request-dependent set of active
ANN queries, all served by a single-vector item index. On Snap DPA, \method{} outperforms four learned multi-interest retrievers on every metric while issuing $33\%$ fewer ANN queries per request. Deployed as a new retrieval source in the production stack, \method{} lifts CVR by $3.11\%$ over the existing retrieval mix and delivers $44\%$ higher CTR and $51\%$ higher CVR than the item-to-item source running beside it.

\bibliographystyle{ACM-Reference-Format}
\bibliography{main}

@inproceedings{li2019mind,
  title     = {Multi-Interest Network with Dynamic Routing for Recommendation at {Tmall}},
  author    = {Li, Chao and Liu, Zhiyuan and Wu, Mengmeng and Xu, Yuchi and Huang, Pipei and Zhao, Huan and Kang, Guoliang and Chen, Qiwei and Li, Wei and Lee, Dik Lun},
  booktitle = {Proceedings of the 28th ACM International Conference on Information and Knowledge Management (CIKM '19)},
  pages     = {2615--2623},
  year      = {2019},
  publisher = {ACM},
  address   = {New York, NY, USA},
  doi       = {10.1145/3357384.3357814},
  eprint    = {1904.08030},
  archivePrefix = {arXiv},
  primaryClass  = {cs.IR}
}

@inproceedings{cen2020comirec,
  title     = {Controllable Multi-Interest Framework for Recommendation},
  author    = {Cen, Yukuo and Zhang, Jianwei and Zou, Xu and Zhou, Chang and Yang, Hongxia and Tang, Jie},
  booktitle = {Proceedings of the 26th ACM SIGKDD International Conference on Knowledge Discovery and Data Mining (KDD '20)},
  pages     = {2942--2951},
  year      = {2020},
  publisher = {ACM},
  address   = {New York, NY, USA},
  doi       = {10.1145/3394486.3403344},
  eprint    = {2005.09347},
  archivePrefix = {arXiv},
  primaryClass  = {cs.IR}
}

@inproceedings{tan2021sine,
  title     = {Sparse-Interest Network for Sequential Recommendation},
  author    = {Tan, Qiaoyu and Zhang, Jianwei and Yao, Jiangchao and Liu, Ninghao and Zhou, Jingren and Yang, Hongxia and Hu, Xia},
  booktitle = {Proceedings of the 14th ACM International Conference on Web Search and Data Mining (WSDM '21)},
  pages     = {598--606},
  year      = {2021},
  publisher = {ACM},
  address   = {New York, NY, USA},
  doi       = {10.1145/3437963.3441811},
  eprint    = {2102.09267},
  archivePrefix = {arXiv},
  primaryClass  = {cs.IR}
}

@inproceedings{xu2022mvke,
  title     = {Mixture of Virtual-Kernel Experts for Multi-Objective User Profile Modeling},
  author    = {Xu, Zhenhui and Zhao, Meng and Liu, Liqun and Xiao, Lei and Zhang, Xiaopeng and Zhang, Bifeng},
  booktitle = {Proceedings of the 28th ACM SIGKDD Conference on Knowledge Discovery and Data Mining (KDD '22)},
  pages     = {4257--4267},
  year      = {2022},
  publisher = {ACM},
  address   = {New York, NY, USA},
  doi       = {10.1145/3534678.3539062},
  eprint    = {2106.07356},
  archivePrefix = {arXiv},
  primaryClass  = {cs.IR}
}

@inproceedings{fan2025dcm,
  title     = {Synergizing Implicit and Explicit User Interests: A Multi-Embedding Retrieval Framework at {Pinterest}},
  author    = {Fan, Zhibo and Lin, Hongtao and Chen, Haoyu and Deng, Bowen and Xia, Hedi and Yan, Yuke and Li, James},
  booktitle = {Proceedings of the 31st ACM SIGKDD Conference on Knowledge Discovery and Data Mining V.2 (KDD '25)},
  pages     = {4396--4405},
  year      = {2025},
  publisher = {ACM},
  address   = {New York, NY, USA},
  doi       = {10.1145/3711896.3737265},
  eprint    = {2506.23060},
  archivePrefix = {arXiv},
  primaryClass  = {cs.IR}
}

@inproceedings{covington2016youtubednn,
  title     = {Deep Neural Networks for {YouTube} Recommendations},
  author    = {Covington, Paul and Adams, Jay and Sargin, Emre},
  booktitle = {Proceedings of the 10th ACM Conference on Recommender Systems (RecSys '16)},
  pages     = {191--198},
  year      = {2016},
  publisher = {ACM},
  address   = {New York, NY, USA},
  doi       = {10.1145/2959100.2959190}
}

@inproceedings{pal2020pinnersage,
  title     = {{PinnerSage}: Multi-Modal User Embedding Framework for Recommendations at {Pinterest}},
  author    = {Pal, Aditya and Eksombatchai, Chantat and Zhou, Yitong and Zhao, Bo and Rosenberg, Charles and Leskovec, Jure},
  booktitle = {Proceedings of the 26th ACM SIGKDD International Conference on Knowledge Discovery \& Data Mining (KDD '20)},
  pages     = {2311--2320},
  year      = {2020},
  publisher = {ACM},
  address   = {New York, NY, USA},
  doi       = {10.1145/3394486.3403280},
  eprint    = {2007.03634},
  archivePrefix = {arXiv},
  primaryClass  = {cs.LG}
}

@inproceedings{huang2013dssm,
  title     = {Learning Deep Structured Semantic Models for Web Search using Clickthrough Data},
  author    = {Huang, Po-Sen and He, Xiaodong and Gao, Jianfeng and Deng, Li and Acero, Alex and Heck, Larry P.},
  booktitle = {Proceedings of the 22nd ACM International Conference on Information \& Knowledge Management (CIKM '13)},
  pages     = {2333--2338},
  year      = {2013},
  publisher = {ACM},
  address   = {New York, NY, USA},
  doi       = {10.1145/2505515.2505665}
}

@inproceedings{yi2019sampling,
  title     = {Sampling-Bias-Corrected Neural Modeling for Large Corpus Item Recommendations},
  author    = {Yi, Xinyang and Yang, Ji and Hong, Lichan and Cheng, Derek Zhiyuan and Heldt, Lukasz and Kumthekar, Aditee and Zhao, Zhe and Wei, Li and Chi, Ed},
  booktitle = {Proceedings of the 13th ACM Conference on Recommender Systems (RecSys '19)},
  pages     = {269--277},
  year      = {2019},
  publisher = {ACM},
  address   = {New York, NY, USA},
  doi       = {10.1145/3298689.3346996}
}

@inproceedings{huang2020facebook,
  title     = {Embedding-based Retrieval in {Facebook} Search},
  author    = {Huang, Jui-Ting and Sharma, Ashish and Sun, Shuying and Xia, Li and Zhang, David and Pronin, Philip and Padmanabhan, Janani and Ottaviano, Giuseppe and Yang, Linjun},
  booktitle = {Proceedings of the 26th ACM SIGKDD Conference on Knowledge Discovery and Data Mining (KDD '20)},
  pages     = {2553--2561},
  year      = {2020},
  publisher = {ACM},
  address   = {New York, NY, USA},
  doi       = {10.1145/3394486.3403305},
  eprint    = {2006.11632},
  archivePrefix = {arXiv},
  primaryClass  = {cs.IR}
}

@inproceedings{vaswani2017attention,
  title     = {Attention Is All You Need},
  author    = {Vaswani, Ashish and Shazeer, Noam and Parmar, Niki and Uszkoreit, Jakob and Jones, Llion and Gomez, Aidan N. and Kaiser, {\L}ukasz and Polosukhin, Illia},
  booktitle = {Advances in Neural Information Processing Systems 30 (NeurIPS)},
  pages     = {5998--6008},
  year      = {2017},
  publisher = {Curran Associates, Inc.},
  address   = {Red Hook, NY, USA},
  eprint    = {1706.03762},
  archivePrefix = {arXiv},
  primaryClass  = {cs.CL}
}

@inproceedings{li2023blip2,
  title     = {{BLIP-2}: Bootstrapping Language-Image Pre-training with Frozen Image Encoders and Large Language Models},
  author    = {Li, Junnan and Li, Dongxu and Savarese, Silvio and Hoi, Steven},
  booktitle = {Proceedings of the 40th International Conference on Machine Learning (ICML)},
  series    = {Proceedings of Machine Learning Research},
  volume    = {202},
  pages     = {19730--19742},
  year      = {2023},
  publisher = {PMLR},
  address   = {Honolulu, HI, USA},
  eprint    = {2301.12597},
  archivePrefix = {arXiv},
  primaryClass  = {cs.CV}
}

@inproceedings{jaegle2022perceiverio,
  title     = {Perceiver {IO}: A General Architecture for Structured Inputs and Outputs},
  author    = {Jaegle, Andrew and Borgeaud, Sebastian and Alayrac, Jean-Baptiste and Doersch, Carl and Ionescu, Catalin and Ding, David and Koppula, Skanda and Zoran, Daniel and Brock, Andrew and Shelhamer, Evan and H{\'e}naff, Olivier J. and Botvinick, Matthew M. and Zisserman, Andrew and Vinyals, Oriol and Carreira, Jo{\~a}o},
  booktitle = {International Conference on Learning Representations (ICLR)},
  numpages  = {29},
  year      = {2022},
  publisher = {OpenReview.net},
  address   = {Virtual Event},
  eprint    = {2107.14795},
  archivePrefix = {arXiv},
  primaryClass  = {cs.LG}
}

@inproceedings{carion2020detr,
  title     = {End-to-End Object Detection with Transformers},
  author    = {Carion, Nicolas and Massa, Francisco and Synnaeve, Gabriel and Usunier, Nicolas and Kirillov, Alexander and Zagoruyko, Sergey},
  booktitle = {Computer Vision -- ECCV 2020},
  series    = {Lecture Notes in Computer Science},
  volume    = {12346},
  pages     = {213--229},
  year      = {2020},
  publisher = {Springer International Publishing},
  address   = {Cham},
  doi       = {10.1007/978-3-030-58452-8_13},
  eprint    = {2005.12872},
  archivePrefix = {arXiv},
  primaryClass  = {cs.CV}
}

@inproceedings{lee2019set,
  title     = {Set Transformer: A Framework for Attention-based Permutation-Invariant Neural Networks},
  author    = {Lee, Juho and Lee, Yoonho and Kim, Jungtaek and Kosiorek, Adam R. and Choi, Seungjin and Teh, Yee Whye},
  booktitle = {Proceedings of the 36th International Conference on Machine Learning (ICML)},
  series    = {Proceedings of Machine Learning Research},
  volume    = {97},
  pages     = {3744--3753},
  year      = {2019},
  publisher = {PMLR},
  address   = {Long Beach, CA, USA},
  eprint    = {1810.00825},
  archivePrefix = {arXiv},
  primaryClass  = {cs.LG}
}

@inproceedings{zhang2022re4,
  title     = {{Re4}: Learning to Re-contrast, Re-attend, Re-construct for Multi-interest Recommendation},
  author    = {Zhang, Shengyu and Yang, Lingxiao and Yao, Dong and Lu, Yujie and Feng, Fuli and Zhao, Zhou and Chua, Tat-Seng and Wu, Fei},
  booktitle = {Proceedings of the ACM Web Conference 2022 (WWW '22)},
  pages     = {2216--2226},
  year      = {2022},
  publisher = {ACM},
  address   = {New York, NY, USA},
  doi       = {10.1145/3485447.3512094},
  eprint    = {2208.08011},
  archivePrefix = {arXiv},
  primaryClass  = {cs.IR}
}

@inproceedings{xie2023remi,
  title     = {Rethinking Multi-Interest Learning for Candidate Matching in Recommender Systems},
  author    = {Xie, Yueqi and Gao, Jingqi and Zhou, Peilin and Ye, Qichen and Hua, Yining and Kim, Jae Boum and Wu, Fangzhao and Kim, Sunghun},
  booktitle = {Proceedings of the 17th ACM Conference on Recommender Systems (RecSys '23)},
  pages     = {283--293},
  year      = {2023},
  publisher = {ACM},
  address   = {New York, NY, USA},
  doi       = {10.1145/3604915.3608766},
  eprint    = {2302.14532},
  archivePrefix = {arXiv},
  primaryClass  = {cs.IR}
}

@article{li2025misurvey,
  title     = {Multi-Interest Recommendation: A Survey},
  author    = {Li, Zihao and Chen, Qiang and Zou, Lixin and Sun, Aixin and Li, Chenliang},
  journal   = {arXiv preprint arXiv:2506.15284},
  year      = {2025},
  eprint    = {2506.15284},
  archivePrefix = {arXiv},
  primaryClass  = {cs.IR}
}

@article{liu2024kuaiformer,
  title     = {{KuaiFormer}: Transformer-Based Retrieval at Kuaishou},
  author    = {Liu, Chi and Cao, Jiangxia and Huang, Rui and Zheng, Kai and Luo, Qiang and Gai, Kun and Zhou, Guorui},
  journal   = {arXiv preprint arXiv:2411.10057},
  year      = {2024},
  eprint    = {2411.10057},
  archivePrefix = {arXiv},
  primaryClass  = {cs.IR}
}

@inproceedings{khattab2020colbert,
  title     = {{ColBERT}: Efficient and Effective Passage Search via Contextualized Late Interaction over {BERT}},
  author    = {Khattab, Omar and Zaharia, Matei},
  booktitle = {Proceedings of the 43rd International ACM SIGIR Conference on Research and Development in Information Retrieval (SIGIR '20)},
  pages     = {39--48},
  year      = {2020},
  publisher = {ACM},
  address   = {New York, NY, USA},
  doi       = {10.1145/3397271.3401075},
  eprint    = {2004.12832},
  archivePrefix = {arXiv},
  primaryClass  = {cs.IR}
}

@inproceedings{santhanam2022colbertv2,
  title     = {{ColBERTv2}: Effective and Efficient Retrieval via Lightweight Late Interaction},
  author    = {Santhanam, Keshav and Khattab, Omar and Saad-Falcon, Jon and Potts, Christopher and Zaharia, Matei},
  booktitle = {Proceedings of the 2022 Conference of the North American Chapter of the Association for Computational Linguistics: Human Language Technologies (NAACL-HLT)},
  pages     = {3715--3734},
  year      = {2022},
  publisher = {Association for Computational Linguistics},
  address   = {Stroudsburg, PA, USA},
  doi       = {10.18653/v1/2022.naacl-main.272},
  eprint    = {2112.01488},
  archivePrefix = {arXiv},
  primaryClass  = {cs.IR}
}

@inproceedings{guo2020scann,
  title     = {Accelerating Large-Scale Inference with Anisotropic Vector Quantization},
  author    = {Guo, Ruiqi and Sun, Philip and Lindgren, Erik and Geng, Quan and Simcha, David and Chern, Felix and Kumar, Sanjiv},
  booktitle = {Proceedings of the 37th International Conference on Machine Learning (ICML)},
  series    = {Proceedings of Machine Learning Research},
  volume    = {119},
  pages     = {3887--3896},
  year      = {2020},
  publisher = {PMLR},
  address   = {Virtual Event},
  eprint    = {1908.10396},
  archivePrefix = {arXiv},
  primaryClass  = {cs.LG}
}

@article{johnson2021faiss,
  title     = {Billion-Scale Similarity Search with {GPUs}},
  author    = {Johnson, Jeff and Douze, Matthijs and J{\'e}gou, Herv{\'e}},
  journal   = {IEEE Transactions on Big Data},
  volume    = {7},
  number    = {3},
  pages     = {535--547},
  year      = {2021},
  doi       = {10.1109/TBDATA.2019.2921572},
  eprint    = {1702.08734},
  archivePrefix = {arXiv},
  primaryClass  = {cs.CV}
}

@inproceedings{jang2017gumbel,
  title     = {Categorical Reparameterization with {Gumbel-Softmax}},
  author    = {Jang, Eric and Gu, Shixiang and Poole, Ben},
  booktitle = {International Conference on Learning Representations (ICLR)},
  numpages  = {12},
  year      = {2017},
  publisher = {OpenReview.net},
  address   = {Toulon, France},
  eprint    = {1611.01144},
  archivePrefix = {arXiv},
  primaryClass  = {cs.LG}
}

@inproceedings{sabour2017capsules,
  title     = {Dynamic Routing Between Capsules},
  author    = {Sabour, Sara and Frosst, Nicholas and Hinton, Geoffrey E.},
  booktitle = {Advances in Neural Information Processing Systems 30 (NeurIPS)},
  pages     = {3856--3866},
  year      = {2017},
  publisher = {Curran Associates, Inc.},
  address   = {Red Hook, NY, USA},
  eprint    = {1710.09829},
  archivePrefix = {arXiv},
  primaryClass  = {cs.CV}
}

@article{kuhn1955hungarian,
  title     = {The {Hungarian} Method for the Assignment Problem},
  author    = {Kuhn, Harold W.},
  journal   = {Naval Research Logistics Quarterly},
  volume    = {2},
  number    = {1--2},
  pages     = {83--97},
  year      = {1955},
  publisher = {Wiley},
  doi       = {10.1002/nav.3800020109}
}

@inproceedings{loshchilov2019adamw,
  title     = {Decoupled Weight Decay Regularization},
  author    = {Loshchilov, Ilya and Hutter, Frank},
  booktitle = {International Conference on Learning Representations (ICLR)},
  numpages  = {19},
  year      = {2019},
  publisher = {OpenReview.net},
  address   = {New Orleans, LA, USA},
  eprint    = {1711.05101},
  archivePrefix = {arXiv},
  primaryClass  = {cs.LG}
}

@article{oord2018cpc,
  title     = {Representation Learning with Contrastive Predictive Coding},
  author    = {van den Oord, A{\"a}ron and Li, Yazhe and Vinyals, Oriol},
  journal   = {arXiv preprint arXiv:1807.03748},
  year      = {2018},
  eprint    = {1807.03748},
  archivePrefix = {arXiv},
  primaryClass  = {cs.LG}
}

\end{document}